# Acetone as a Polar Cosolvent for Pyridinium-Based Ionic Liquids

Vitaly V. Chaban

Federal University of São Paulo, São Paulo, Brazil

**Abstract**. The work reports transport and structure properties of the three pyridinium-based ionic liquids (RTILs) — N-butylpyridinium hexafluorophosphate [BPY][PF$_6$], N-butylpyridinium trifluorosulfonate [BPY][TF] and N-butylpyridinium bis(trifluoromethanesulfonyl)imide [BPY][TFSI] — in their mixtures with acetone (ACET). The ionic conductivity maximum occurs at 10 mol% RTIL, irrespective of the anion. The absolute ionic conductivity value of [BPY][TF] is higher than those of other RTILs. This is explained by a weaker cation-anion binding than that in [BPY][PF$_6$] and smaller anion size than that of [BPY][TFSI]. All the investigated RTILs are infinitely miscible with ACET, which boosts their diffusivity and conductivity. A small viscosity of ACET favors a drastic viscosity decay in the RTIL-ACET mixtures. Structure analysis (radial distribution functions, ionic clusters) of the mixtures are in line with their transport properties providing a reliable microscopic interpretation of the observed macroscopic properties. As a molecular co-solvent, ACET constitutes an interesting alternative and competitor to more intensively investigated liquids.

TOC Image

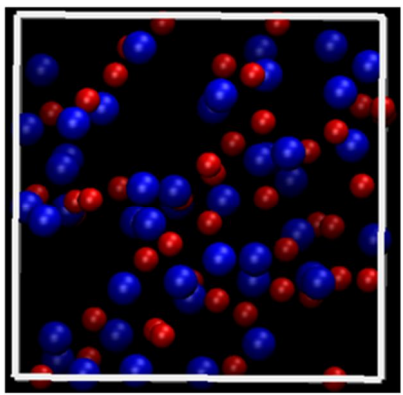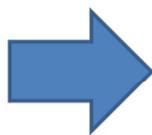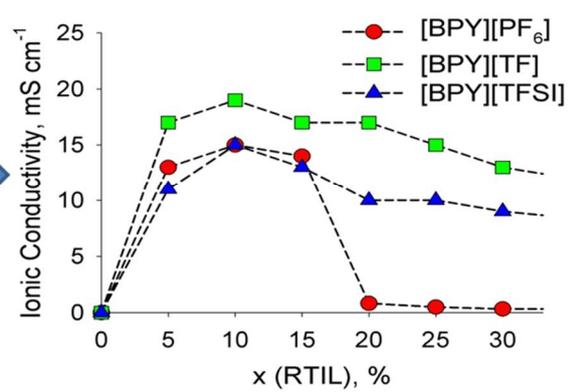

**Introduction**

Room-temperature ionic liquids (RTILs) constitute an actively developing field thanks to their versatility and an interesting set of physical chemical properties.[1-7] Tunability of RTILs due to the many available functionalization pathways and readily combined organic ions make them 'designer solvents'. Ideally, a virtually continuous range of physical chemical properties can be achieved to target each particular application. RTILs exhibit wide electrochemical windows,[8,9] relatively high ionic conductivities,[8,10] low volatilities,[5,11] excellent solvation and gas adsorption properties.[12,13] Amino functionalized ILs provide an efficient carbon dioxide capture opportunities through chemisorption,[14] behaving in a similar way to aqueous amine solutions. Attaching different functional groups to the same cation or anion opens a bright avenue to develop ever better scavengers and more universal solvents.

Ionic conductivity of RTILs can be further increased by adding a polar molecular co-solvent, such as water, alcohols, acetonitrile, ketones, etc.[10,15] A dependence of ionic conductivity on solvent molar fraction exhibits a maximum,[16-18] which is located below 20 mol % RTIL, irrespective of the particular ionic liquid. Therefore, a significant amount of a molecular admixture is required to immobilize charge carriers. Shear viscosity of the resulting ion-molecular mixtures exponentially (vs. molar fraction) decays upon dilution. An absolute value of the conductivity maximum is determined by (1) diffusivity of a co-solvent; (2) sizes and shapes of ions; (3) cation-anion binding; (4) IL-co-solvent binding. The two latter contributions appear simultaneously relevant for the homogeneity of the mixtures. A significant collection of experimental data is available[10,19,20] confirming that smaller ions and mobile co-solvent molecules favor a higher ionic conductivity, whereas bulky ions and relatively (e.g. dimethyl sulfoxide and propylene carbonate) result in lower values. It is important for manifold electrochemical applications of ILs (solar cells, supercapacitors, lithium-ion batteries)[21-24] to have a properly conductive electrolyte.

While aqueous, alcoholic and acetonitrile solutions/mixtures of popular RTILs have been addressed in the past, acetone (ACET) seems somewhat underestimated as a co-solvent. Relatively few works are available devoted to the imidazolium-based RTILs in ACET and some other RTILs sporadically.[25-37] Based on its molecular and physical chemical properties, ACET excellently mixes with most ILs thanks to the carbonyl group. Furthermore, ACET exhibits a low viscosity, 0.307 cP,[38] which correlates with its low normal-pressure boiling point. Unlike most other solvents, the dipole moment of ACET is perpendicular to its backbone. This feature is expected to give rise to a different local structure pattern. ACET does not possess a suitable hydrogen atom to get involved into hydrogen bonding. Therefore, diffusion of pure ACET is notably high, while viscosity is notably low, despite a relatively big molecular mass.

In the present work, we report a detailed investigation of the ion-molecular mixtures consisting of the three N-butylpyridinium-based ILs and acetone (Figure 1). N-butylpyridinium hexafluorophosphate, N-butylpyridinium trifluoromethanesulfonate, N-butylpyridinium bis(trifluoromethanesulfonyl)imide RTILs were chosen in view of the insufficient researchers' attention to them thus far. In view of the expected location of the conductivity maximum, we concentrate our attention on the 5-30 mol% RTIL mixtures. Higher concentrations of RTILs are omitted. However, these concentrations may constitute interest for solvation applications, which are outside our present scope. Detailed analysis of structure and transport properties is provided based on the atomistically-precise pairwise-potential-based molecular dynamics (MD) simulations. The MD simulations were carried out by means of our recently developed force field for the pyridinium family of RTILs.[20]

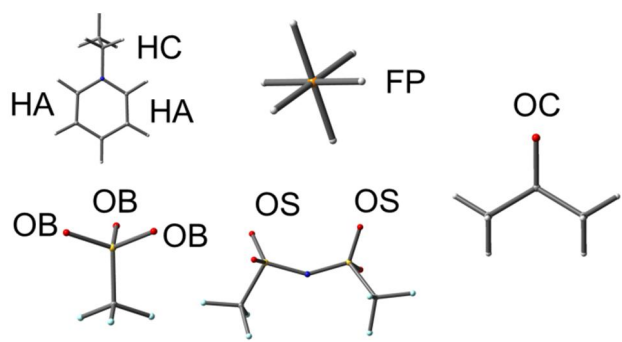

Figure 1. Optimized geometries in vacuum of the simulated molecules and ions: N-butylpyridinium cation, hexafluorophosphate anion, trifluoromethanesulfonate anion; bis(trifluoromethanesulfonyl)imide anion, and acetone molecule. The selected interaction sites — HA, HC, FP, OB, OS, OC — are shown. These sites will be used to describe ion-molecular structures in the liquid phase.

**Simulation Details**

Pairwise-potential-based MD simulation method is a powerful computer simulation tool that is widely employed nowadays to describe systems containing thousands to millions atoms. If parametrization of the interaction potentials is performed carefully, the results provide highly reliable data, whose discrepancies from the experimental data rarely exceed 10%.

The force field (FF) for the pyridinium family of RTILs was developed by our group recently.[20] This FF accurately reproduces diffusivity, conductivity, viscosity and thermodynamic quantities of the pure pyridinium-based RTILs at room and elevated temperatures. The ACET molecule was represented using the classical model proposed by Jorgensen and coworkers.[39] Both models originate from the OPLS-AA methodology.[40] Thus, these models are fully compatible.

Table 1 provides an overview of the studied MD systems, whereas Figure 2 exemplifies equilibrated model cells of the corresponding RTILs and ACET. N-butylpyridinium hexafluorophosphate, [BPY][PF$_6$], N-butylpyridinium trifluoromethanesulfonate [BPY][TF], and N-butylpyridinium bis(trifluoromethanesulfonyl)imide [BPY][TFSI] were randomly immersed into cubic cells containing the pre-equilibrated ACET molecules. The correct density of each

MD system was achieved using the Parrinello-Rahman barostat.[41] The reference pressure of the barostat was 1.0 bar; the relaxation constant was 1.0 ps; the isobaric compressibility constant was 4.5 $10^{-5}$ bar$^{-1}$. The constant temperature, 300 K, was maintained using the Bussi-Donadio-Parrinello thermostat[42] with a relaxation constant of 1.0 ps. The equilibration time for each system was 10 000 ps. The equations-of-motion of each atom, including hydrogen atoms, were propagated with a time-step of 0.002 ps. The sampling times (Table 1) were selected according to the expected viscosity of each MD system. The atomistic trajectories were propagated by GROMACS 4.[43]

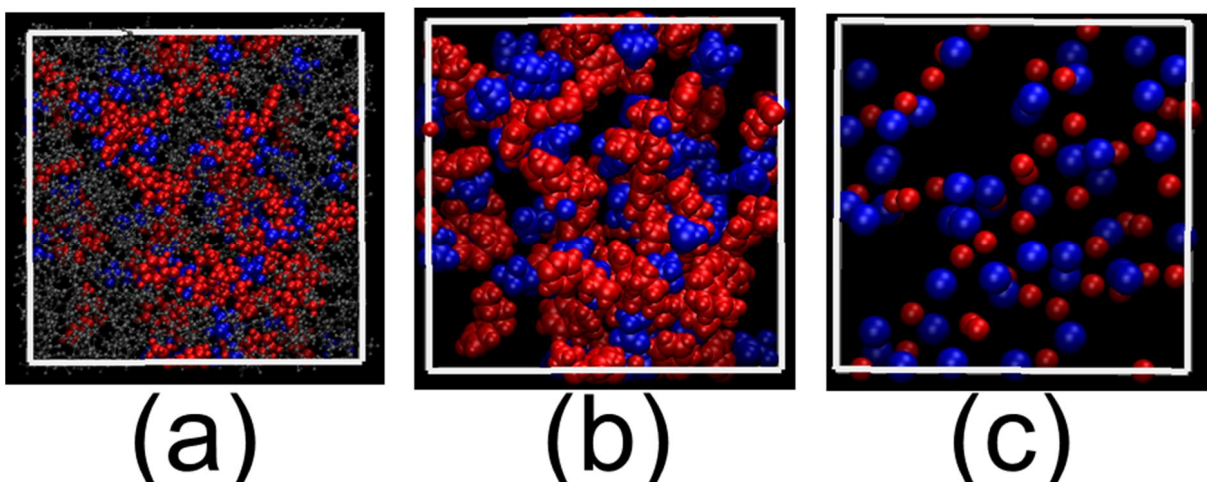

Figure 2. An equilibrated molecular dynamics box: (a) all ions and molecules; (b) cations and anions; (c) centers-of-mass of cations and anions. The cations are red, the anions are blue, the solvent molecules are grey. Spheres of ions were enlarged twice for better visibility. The system contains 45 [BPY][TF] ion pairs and 405 ACET molecules corresponding altogether to 5490 interaction centers. The images were prepared in VMD 1.9.[44]

Table 1. Simulated compositions of ion-molecular systems containing the pyridinium-based RTILs and ACET. The more dilute MD systems can be sampled for a somewhat shorter time, since ionic and molecular dynamics in these MD systems is faster than in the more concentrated MD systems. Transport of particles increases exponentially upon dilution (expressed as molar fraction)

| # | N (RTIL) | N (acetone) | N (atoms) | Molar fraction, % | Sampling time, ns |
|---|---|---|---|---|---|
| N-butylpyridinium hexafluorophosphate [BPY][PF$_6$] + acetone | | | | | |
| 1 | 25 | 475 | 5525 | 5 | 50 |
| 2 | 45 | 405 | 5445 | 10 | 50 |
| 3 | 60 | 340 | 5260 | 15 | 50 |
| 4 | 75 | 300 | 5325 | 20 | 70 |
| 5 | 90 | 270 | 5490 | 25 | 80 |
| 6 | 105 | 245 | 5705 | 30 | 100 |
| N-butylpyridinium trifluoromethanesulfonate [BPY][TF] + acetone | | | | | |
| 7 | 25 | 475 | 5550 | 5 | 50 |
| 8 | 45 | 405 | 5490 | 10 | 50 |
| 9 | 60 | 340 | 5320 | 15 | 50 |
| 10 | 75 | 300 | 5400 | 20 | 70 |
| 11 | 86 | 258 | 5332 | 25 | 70 |
| 12 | 96 | 224 | 5312 | 30 | 80 |
| N-butylpyridinium bis(trifluoromethanesulfonyl)imide [BPY][TFSI] + acetone | | | | | |
| 13 | 25 | 475 | 5725 | 5 | 50 |
| 14 | 40 | 360 | 5160 | 10 | 50 |
| 15 | 60 | 340 | 5740 | 15 | 50 |
| 16 | 65 | 260 | 5135 | 20 | 60 |
| 17 | 75 | 225 | 5175 | 25 | 60 |
| 18 | 90 | 210 | 5610 | 30 | 70 |

We simulated electrostatic energy between each two charged interaction centers belonging to molecules and ions directly using the Coulomb law if the distance between the two interaction centers did not exceed 1.2 nm. In the case of larger distances, the particle-mesh Ewald method was used.[45] Lennard-Jones (12,6) forces were modified between 1.1 and 1.2 nm using the shifted force scheme. An accurate treatment of the interaction potentials in combination with a sufficiently small trajectory propagation time-step preserves total energy of the MD system. Intra-molecular interactions were reproduced using harmonic potentials for bonds, covalent angles, and covalent dihedrals. The Coulombic and Lennard-Jones (12,6) potentials for atoms separated by one, two, and three bonds were omitted, whereas 1-4 interactions were decreased twice. This is in accordance with the OPLS-AA FF.[40]

Radial distribution functions (RDFs) were computed using their classical definition, in which height of RDF is a ratio of local density to average density of a given atom. Self-diffusion coefficients and ionic conductivity were calculated using the Einstein-type equations for these properties, i.e. based on the mean-squared displacements of atoms. Shear viscosity was calculated based on the energy dissipation.[46] An ionic cluster was defined as a stable collection of ions, which exhibit strong electrostatic binding to one another. The clusterization threshold was set to a position of the first minimum of the HA-FP or HA-OB or HA-OS RDFs, where applicable.

**Results and Discussion**

The ionic conductivity dependence vs. molar fraction (Figure 3) of the selected RTILs is our major research scope. The addition of ACET significantly increases conductivity of all ILs. The conductivity maximum occurs at 10 mol% RTIL, irrespective of the RTIL identity. The conductivity of [BPY][TF] is somewhat larger than those of [BPY][TFSI] and [BPY][PF$_6$], 19 vs. 15 mS cm$^{-1}$ in the point of maximum. The conductivity of [BPY][PF$_6$] decreases drastically upon the ACET content decrease. In turn, conductivity of other RTILs decays more smoothly. This observation may indicate formation of the [BPY]-[PF$_6$] aggregates above certain concentration threshold (15 mol% IL). Viscosity of the ion-molecular mixtures (Figure 4) appears in concordance with ionic conductivity. A high viscosity of the [BPY][PF$_6$] mixtures is explained by a strong cation-anion binding. The [TF] and [TFSI] anions are similar chemically, therefore smaller viscosity of the [BPY][TF] containing mixtures is explained by a smaller size of the corresponding anion.

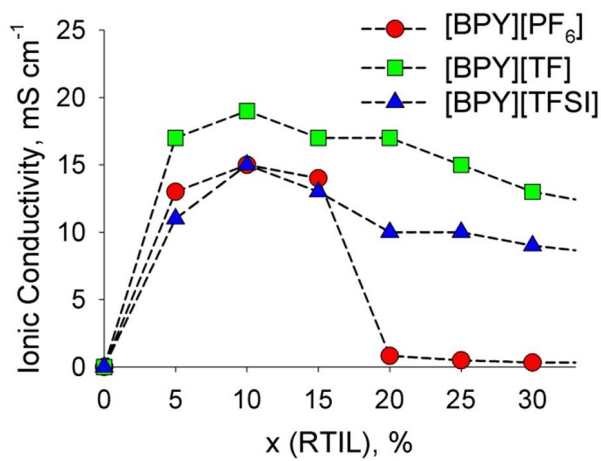

Figure 3. Ionic conductivity vs. molar fraction of RTIL in the three pyridinium-based RTILs at 300 K and 1 bar. Acetone is a co-solvent.

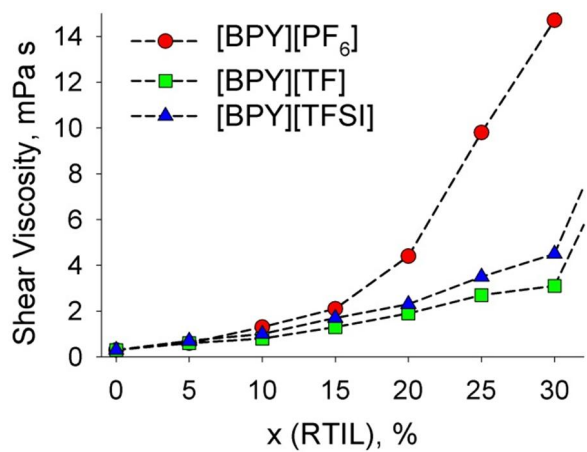

Figure 4. Shear viscosity vs. molar fraction of RTIL in the three pyridinium-based RTILs at 300 K and 1 bar. Acetone is a co-solvent. Shear viscosity of pure acetone amounts to 0.3 mPa s.

Addition of a molecular co-solvent, such as ACET, allows to exponentially increase diffusivity of ions (Figure 5). Even though concentration of charge carriers (ions) decreases upon dilution, their higher mobility brings an ionic conductivity increase until the concentration becomes smaller than 10 mol% RTIL in all cases. Interestingly, diffusion of ACET is very small in the more concentrated [BPY][PF$_6$] containing mixtures, as compared to that in other RTILs. Anions are systematically faster than the N-butylpyridinium cation. The exception is [TFSI] at higher RTIL contents. Higher diffusivity of other anions is commensurate with the corresponding anion size and mass differences.

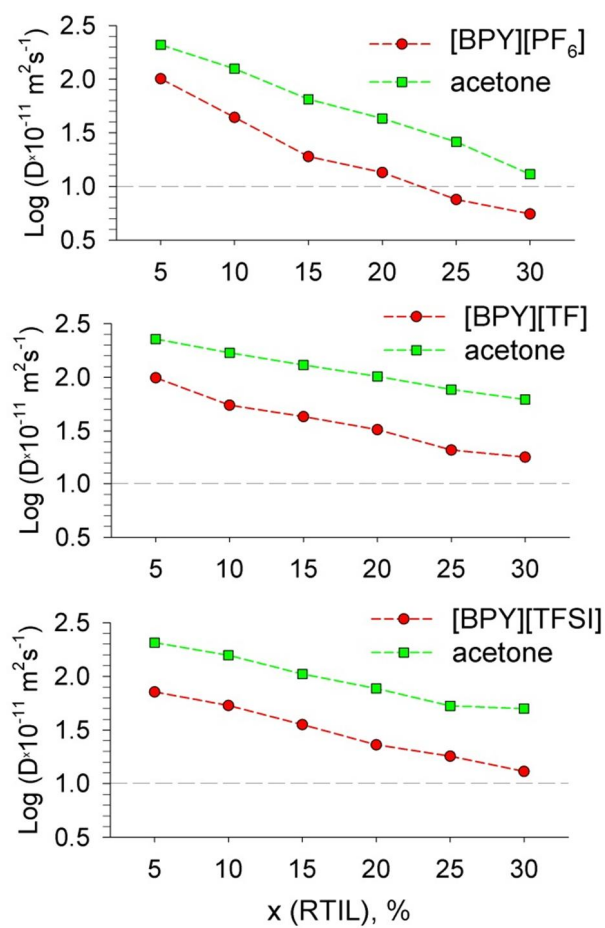

Figure 5. Self-diffusion coefficients of the three pyridinium-based RTILs, $(D_{cation} + D_{anion}) / 2$, and ACET molecules vs. molar fraction of RTIL at 300 K and 1 bar. The long-dashed grey line, $\log (D) = 1$, is provided for simpler comparison of data. The identical scale of all plots allows an easier comparison of the ACET solvation effect.

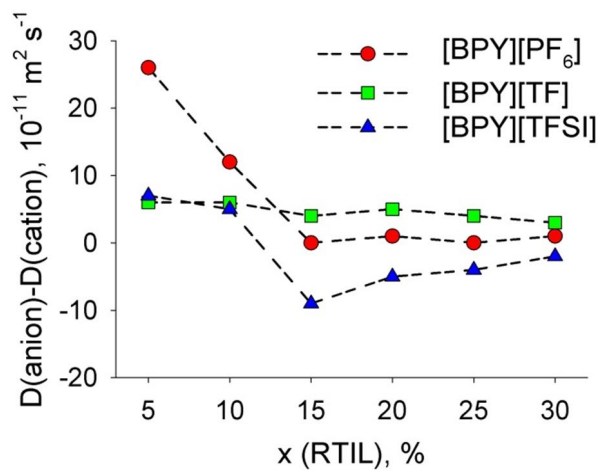

Figure 6. Difference between diffusion coefficients of anions and cations in three pyridinium-based liquids vs. molar fraction of RTIL at 300 K and 1 bar. ACET is a co-solvent.

RDFs were computed for certain partially charged interaction sites representing the cation, the anion and the solvent molecules (Figures 7-8) in an attempt to identify strong atom-atom distance correlations. ACET does not form a hydrogen bond (H-bond) with the N-butylpyridinium cation, since all relevant peaks are located at 0.26-0.27 nm. The distance between the hydrogen atom of the pyridine ring and the fluorine atom of the [PF$_6$] anion equals to 0.26 nm, which is also too large to qualify for any type of H-bond. Noteworthy, the distance between the same fluorine atom and the closest hydrogen atom of the butyl chain is unusually small, 0.22 nm. However, the hydrogen atoms in the butyl chain are not sufficiently electron deficient to participate in H-bonding. The same conclusions generally apply to the RDFs in [BPY][TF] and [BPY][TFSI].

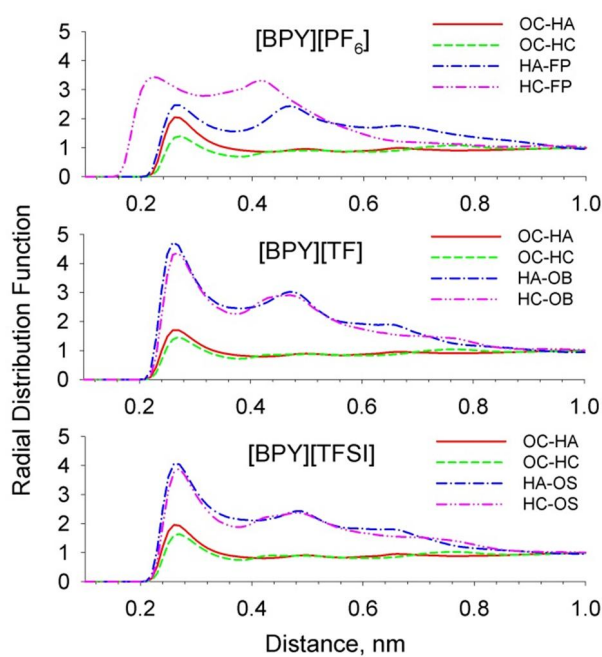

Figure 7. Radial distribution functions for the selected interaction sites of the N-butylpyridinium cation (HA, HC), solvent (OC), and anions (FP, OB, OS). See Figure 1 for an atom type designation. The RDFs were calculated in the 5 mol% RTIL systems and are qualitatively independent on the composition. The analysis of these distributions helps to investigate possible hydrogen bonding in the simulated MD systems. The scale of all plots is made the same for simpler comparison.

In the mixtures, ACET weakly interacts with the [PF$_6$] anion. For instance, the fluorine-oxygen peak is completely smashed. The fluorine-nitrogen peak exhibits a height of 3.5 being located at ca. 0.4 nm. The nitrogen of [BPY] – oxygen of [TF] peak is even higher, 4.5 units. In [BPY][TFSI], nitrogen of [BPY] is separated by 0.4 nm from oxygen of ACET. The cation-anion distance correlations are significantly stronger than all other correlations in all investigated RTILs. The pyridine ring differs from the imidazole ring by its inability to offer strong cation-anion coordination sites. This poor coordination ability also applies to ACET. The size of the [BPY] cations prevents a high diffusivity and, therefore, a high conductivity. Decreasing the grafted hydrocarbon chain, e.g. down to N-methylpyridinium, is expected to boost conductivity significantly.

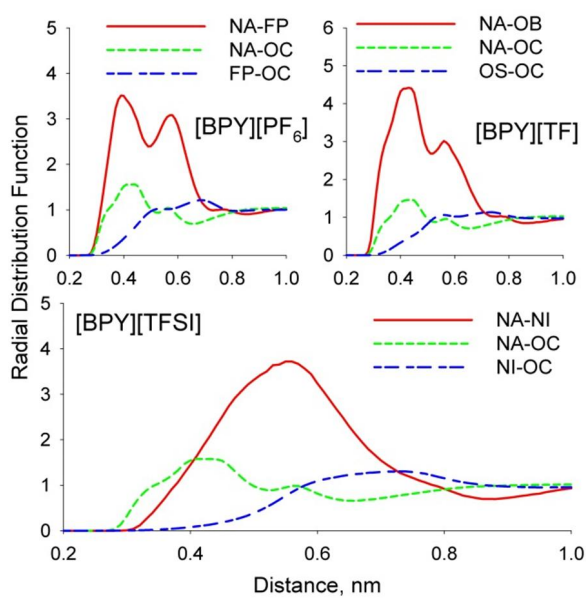

Figure 8. Radial distribution functions for the selected interaction sites of the cation (NA), the three anions (FP, OB, NI), and the solvent (OC). The depicted RDFs provide comparison between solvation of the cation and the anion in ACET and the binding strength within an ion pair. The RDFs were calculated in the 5 mol% RTIL MD systems and are qualitatively independent on the composition.

The percentage of lone ions in the mixtures increases smoothly with the decrease of RTIL molar fraction (Figure 9). The amount of lone ions is systematically smaller in [BPY][TFSI]. In turn, ion pairs exhibit a dependence, which is close to logarithmic and the smallest amount of ion pairs was detected in [BPY][PF$_6$]. Existence of lone ions favors charge mobility, and therefore, conductivity, while an ionic aggregation needs to be suppressed. Higher conductivity of [BPY][TF] in ACET agrees well with Figure 9. In most mixtures, the size of the largest ionic cluster in [BPY][TF] appears smaller than in other RTILs (Figure 10). It must be underlined, however, that the cation-anion binding strength and the RTIL-co-solvent binding strength are also very important factors influencing charge mobility.

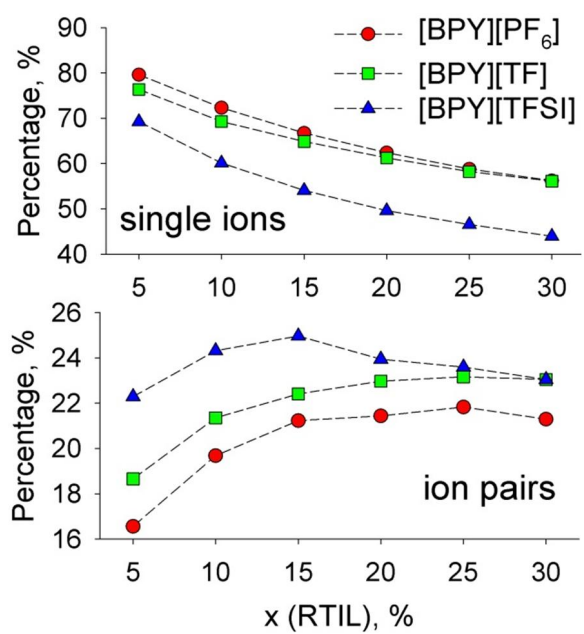

Figure 9. Percentages of lone ions (top) and neutral ion pairs (bottom) in the investigated ion-molecular systems as a function of RTIL molar fraction.

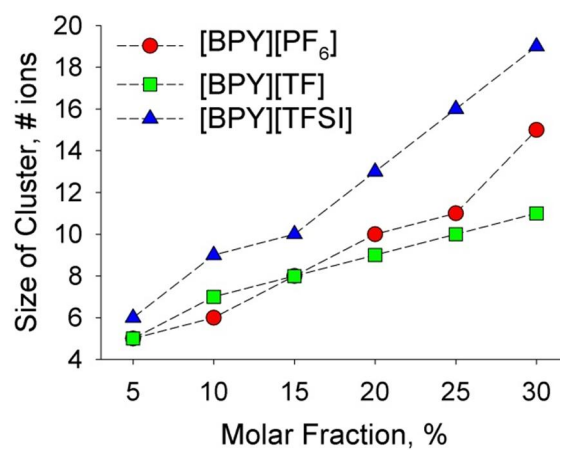

Figure 10. The largest ionic cluster as a function of RTIL molar fraction. Only the clusters with a formation probability of more than 0.1% are included.

Figure 11 summarizes ion-ionic and ion-molecular interaction energies in the RTIL-ACET mixtures. At ca. 20 mol% RTIL, both thermodynamic quantities are similar. This point does not coincide with the conductivity maximum and unlikely can be used for its prediction. Nevertheless, the ratio of interaction energies in the mixtures helps to hypothesize compositions of the first coordination shells of the ions. If an ion is preferentially surrounded by ACET molecules, its diffusivity is higher, as evidenced by the above discussed, Figures 5-6. The highest cation-anion pairing energy is observed in [BPY][TFSI], 76 kJ mol$^{-1}$, whereas the lowest one is in [BPY][TF], 58 kJ mol$^{-1}$. Both values are given at the maximum conductivity composition, 10 mol% RTIL. Ion-ionic energy decreases upon further dilution. The energies depend significantly on the mixture composition. The total energy per the [BPY] cation is also lowest in [BPY][TF], 157 kJ mol$^{-1}$, vs. 174 and 172 kJ mol$^{-1}$ for [BPY][TFSI] and [BPY][PF$_6$], respectively. Significant dependence of the ion pairing energy was observed on the mixture composition. In the 30 mol% RTIL mixture, it amounts to 132 kJ mol$^{-1}$, whereas being twice smaller, 73 kJ mol$^{-1}$, in the 5 mol% RTIL mixture.

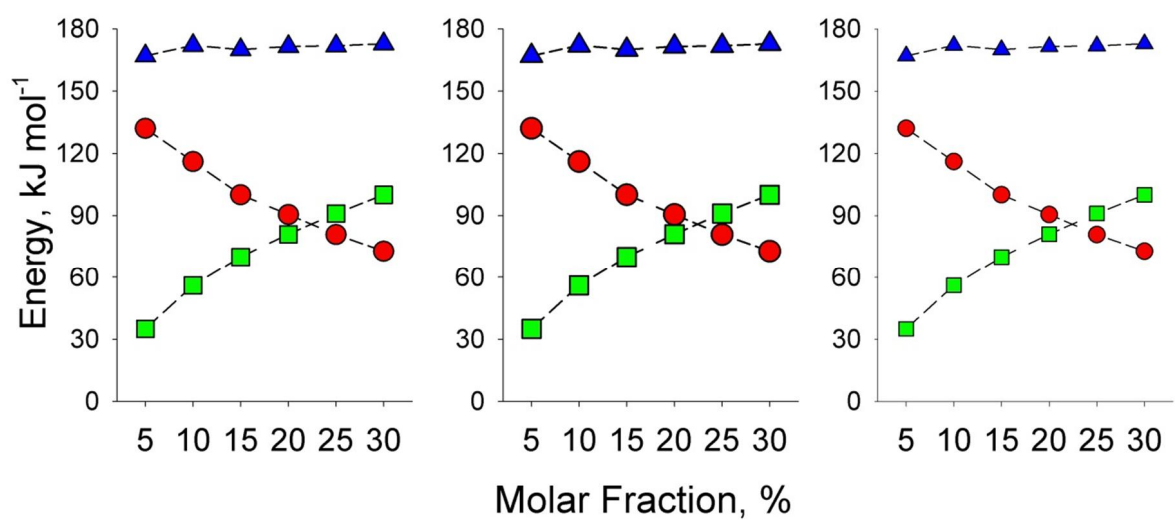

Figure 11. Solvation (red circles), ion-ionic (green squares), and total (blue triangles) pairwise energies per one mole of RTIL in ACET as a function of RTIL molar fraction.

**Conclusions**

Ion-molecular ensembles are important for numerous applications in pure and applied chemistry. Due to their non-ideality, most of such systems cannot be described by simple equations of physical chemistry, nor in many cases of practical interest by means of the continuum theory. Knowledge about ionic conductivity of an electrolyte solution is of key importance for electrochemical applications. This transport property depends on solute concentration, mobility of its particles and ion clustering upon thermal motion. The present work reports comprehensive analysis of the structure, transport and thermodynamic properties in the pyridinium-based RTILs / acetone mixtures at room conditions. Dependence of ionic conductivity upon RTIL molar fraction is explained using shear viscosity, self-diffusion, structure distributions and solvation energies. The highest ionic conductivity was found to occur in the 10 mol% solutions of all RTILs. This is the concentration at which all available ions contribute the resulting conductivity efficiently. Upon addition of ions, conductivity of [BPY][TF] and [BPY][TFSI] decreases insignificantly, whereas conductivity of [BPY][PF$_6$] drops down drastically due to shear viscosity increase and ionic diffusion decrease, accordingly. The content of ion pairs modestly increases as molar fraction of RTIL increases. Nevertheless, this property alone does not explain conductivity trends. The relatively high conductivity of [BPY][TF] is in concordance with a high percentage of single ions in this mixture and a small anion size.

The reported quantities and the established correlations are of direct assistance for future development of binary electrolyte mixtures involving RTILs and polar molecular co-solvents. The pyridinium-based RTILs constitute practical interest due to a relatively weak cation-anion binding and absence of strong hydrogen bonds in these ion-molecular systems.

**Acknowledgments**

V.V.C. is funded through CAPES in Brazil. Iuliia Voroshylova has prepared non-equilibrated GRO files and TOP files for GROMACS.